\documentclass[english]{lni}

\usepackage{amsmath,graphicx}
\usepackage{todonotes}
\usepackage{subfigure}

\def\x{{\mathbf x}}

\begin{document}
\title{Content-based Recommendations for Radio Stations with Deep Learned Audio Fingerprints}

\author[Stefan Langer \and Liza Obermeier \and André Ebert \and Markus Friedrich \and Emma Munisamy]
{
Stefan Langer
\footnote{Ludwig-Maximilians-Universität München, Institut für Informatik, Oettingenstrasse 67, 80538 Munich, Germany,
\email{stefan.langer@ifi.lmu.de}}
\and
Liza Obermeier
\footnote{inovex GmbH, Data Management and Analytics, Lindberghstrasse 3, 80939 Munich,
Germany, 
\email{lobermeier@inovex.de}} 
\and
André Ebert
\footnote{inovex GmbH, Data Management and Analytics, Lindberghstrasse 3, 80939 Munich,
Germany, 
\email{aebert@inovex.de}} 
\and
Markus Friedrich
\footnote{Ludwig-Maximilians-Universität München, Institut für Informatik, Oettingenstrasse 67, 80538 Munich, Germany,
\email{markus.friedrich@ifi.lmu.de}}
\and
Emma Munisamy
\footnote{Ludwig-Maximilians-Universität München, Institut für Informatik, Oettingenstrasse 67, 80538 Munich, Germany,
\email{emma.munisamy@ifi.lmu.de}}
\and
Claudia Linnhoff-Popien
\footnote{Ludwig-Maximilians-Universität München, Institut für Informatik, Oettingenstrasse 67, 80538 Munich, Germany,
\email{linnhoff@ifi.lmu.de}}
}

\editor{Herausgeber et al.} 
\booktitle{Informatik 2020} 
\year{2020}
\maketitle





\begin{abstract}
The world of linear radio broadcasting is characterized by a wide variety of stations and played content.
That is why finding stations playing the preferred content is a tough task for a potential listener, especially due to the overwhelming number of offered choices.
Here, recommender systems usually step in but existing content-based approaches rely on metadata and thus are constrained by the available data quality.
Other approaches leverage user behavior data and thus do not exploit any domain-specific knowledge and are furthermore disadvantageous regarding privacy concerns.
Therefore, we propose a new pipeline for the generation of audio-based radio station fingerprints relying on audio stream crawling and a \textit{Deep Autoencoder}.
We show that the proposed fingerprints are especially useful for characterizing radio stations by their audio content and thus are an excellent representation for meaningful and reliable radio station recommendations.
Furthermore, the proposed modules are part of the \textit{HRADIO Communication Platform}, which enables hybrid radio features to radio stations. 
It is released with a flexible open source license and enables especially small- and medium-sized businesses, to provide customized and high quality radio services to potential listeners.    
\end{abstract}

\begin{keywords}
Hybrid Radio, Multimedia Services, Recommender Systems, Unsupervised Learning, Deep Audio Fingerprints, Deep Learning
\end{keywords}

\section{Introduction}\label{sec:intro}
Despite emerging competition from on-demand content services, linear radio broadcasting still remains one of the most popular entertainment and information media in Europe.
Its advantage lies in its technical simplicity, its topicality, and its personal approach conveyed by professional moderators.
However, with services like Spotify, Deezer, or Google Play, strong competitors have recently appeared.
Those on-demand media services have the advantage of providing a personalized listening experience and a wide range of contents combined with precise recommendation systems.
\par
The challenge for radio broadcasters is now to enrich their classic, linear radio programme with online-based, personalized technologies in order to improve the listening experience and to bridge the gap between linear and on-demand content providers.
So-called hybrid radio technologies comprise of techniques for privacy-preserving user data collection that are capable of providing individual and secure feedback channels. 
These channels enable customers to actively take part in the composition of their radio programme. 
Another important enhancement compared to the established linear programme design is the on-the-fly substitution of content, e.g., replacing ads with songs from a pre-selected music playlist or the skipping of disliked content from one radio station with preferred content from another radio station.
In combination with other hybrid radio technologies, such features help mainstream stations as well as small radio businesses to catch up with online streaming services and to even surpass them in specific disciplines, e.g., interactivity and feedback channel communication.
To realize such rich features, a precise and comprehensive recommender engine is necessary.
Therefore, on the one hand a large amount of well-structured and rich data is needed, on the other hand different concepts for the analysis of meta and audio data have to be selected, implemented and carefully evaluated.
\par
This paper focuses on the latter and thus on the question of how meaningful recommendations for radio stations can be generated on the basis of rich metadata provided by the \textit{HRADIO Communication Platform}, which was implemented by the authors within the scope of previous work \cite{friedrich2019distributed}.
The findings and implemented concepts presented in this work are part of the open accessible HRADIO project\footnote{\url{https://www.hradio.eu/} - The HRADIO project and thus this work was funded by H2020, the EU Framework Programme for Research and Innovation. is driven by several radio stations, small development companies, technology experts, research institutes and the Ludwig-Maximilians-University Munich}.
In the following, two main approach categories are distinguished for analyzing and recommending radio stations and programmes: Recommendations based on \textit{Collaborative Filtering} (CF) and \textit{Content-based Filtering} (CBF) approaches. 
CF focuses on user opinions and behavior, which can lead to privacy issues as well as the \textit{Cold Start Problem} (see Section \ref{sec:relatedwork}). 
Thus, a CBF-based recommender system working with characteristics derived from available station data is preferred in context of this work. 
In \cite{friedrich2019distributed}, it could be shown that station recommendation based on available metadata is possible, but only if the metadata reaches a certain quality level. 
This is provably often not the case. 
Other meaningful characteristics can be directly derived from the station's audio signal which significantly reduces metadata quality requirements.
Therefore, we propose a \textit{Deep Learning}-based audio crawling and fingerprint extraction pipeline for the characterization of radio stations and show visual results for numerous stations. 
Furthermore, we detail on how a recommender system based on the developed station fingerprints can be implemented.
\par
Due to these unique technical circumstances as well as the fact that the \textit{HRADIO Communication Platform} is released under a permissive open source license and available at no cost, new opportunities are offered especially for small- and medium-sized businesses within the radio landscape. It significantly increases the visibility of radio stations and makes them searchable via a search service that can be used by websites and mobile applications regardless of the size of the company. For this purpose, the daily programme and the audio stream of a station are analyzed. This is done automatically, without additional effort and without costs. This creates enormous added value for both radio stations and potential listeners. In particular, small or regional stations that were previously unknown to non-regional listeners can now be found on the basis of a listener's programme taste. In addition, previously untapped potentials are opened up with regard to findability, routing and provision of niche topics as well as the allocation of tailor-made advertising relevant only to certain listener groups.
\par
To present the underlying concept, its implementation, and its evaluation of these ideas, this paper is structured as follows: 
Section \ref{sec:relatedwork} provides a brief overview of recommendation concepts and related work. Section \ref{sec:concept} explains the basic principles of the proposed fingerprinting pipeline and its implementation.
Its recommendation capabilities are evaluated in Section \ref{sec:evaluation} while Section \ref{sec:conclusion} summarizes this work.

\section{Related Work}\label{sec:relatedwork}
Next to recommendation approaches such as \textit{You May Like} (YML), \textit{Knowledge-based Filtering} (KF), and \textit{Demographic Filtering} (DF), there are two mainly recognized concepts: \textit{Content-based Filtering} (CBF) and \textit{Collaborative Filtering} (CF) \cite{bennett2007netflix, burke2002hybrid, ricci2011introduction}.
CF-, DF-, and KF-based systems show good performance if enough user data is available. 
But an open issue is the so-called \textit{Cold Start Problem}, which occurs in the initial phase of the system where not enough user data is available to create meaningful recommendations \cite{chang2015space,lam2008addressing,camacho2018social}. 
Another issue for CF and DF systems are so-called \textit{Filter Bubbles}, describing the creation of closed-off, synthetic environments, in which always the same items are recommended, disregarding the existence of contrary or different items beyond the bubble \cite{pariser2011filter,geschke2019triple}.
In contrast to that, an initial issue of CBF-based systems is the need for high-quality metadata precisely describing the items to recommend \cite{friedrich2019distributed}. 
The concept presented in this paper is part of the \textit{HRADIO Communication Platform}, which provides a vast amount of metadata and audio information within a hybrid radio context \cite{friedrich2019distributed} and is completely open source.
For this reason, and to avoid issues like \textit{Filter Bubbles} or the \textit{Cold Start Problem}, it takes a CBF-based approach.
\par
In order to compare radio services on basis of their audio features, existing approaches which utilize \textit{Deep Learning} for music genre recognition (MGR) can be utilized \cite{dieleman2011audio,oramas2018multimodal}. 
Gwardys et al. propose a concept using transfer learning in combination with a convolutional neural network for MGR \cite{gwardys2014deep}. 
Logan et al. and Siddiquee et al. present methods for measuring the similarity of music on basis of audio signals \cite{logan2001music, siddiquee2016association}. 
After clustering raw audio features, Logan et al. compare entities using the \textit{Earth Movers Distance} (EMD) \cite{rubner2000earth}.
{\c{C}}ataltepe et al. use adaptive features and user grouping to take note to the aspect that different traits within music are of different importance for each user by including historical information about the users' listening behaviour \cite{ccataltepe2007music}. 
Together with others, these works provide valuable input for the proposed concept.

\section{Concept}\label{sec:concept}


\begin{figure}[!htbp]
\centering
\vspace{4mm}
\includegraphics[width=0.8\textwidth]{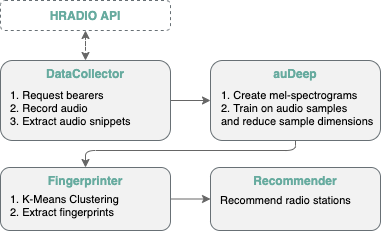}
\vspace{4mm}
\caption{The proposed pipeline consists of four modules: the \textit{Data Collector}, the \textit{auDeep} autoencoder, the \textit{Fingerprinter}, and the \textit{Recommender}. The \textit{Data Collector} ensures the binding to the \textit{HRADIO API}.}
\vspace{1mm}
\label{fig:architecture}	
\end{figure}

This chapter details the steps of the proposed analysis pipeline consisting of a \textit{Data Collector}, the \textit{auDeep} autoencoder, the \textit{Fingerprinter}, and a \textit{Recommender}, as depicted in Figure \ref{fig:architecture}. 
In this context, the \textit{Data Collector} unit (see Chapter \ref{chapter:data-collector}) ensures the binding of the pipeline to the \textit{HRadio API} (which is used for requesting a list of radio services from the \textit{HRADIO Communication Platform} \cite{friedrich2019distributed}), records their audio streams and extracts audio bytes in predefined intervals. The raw audio snippets are transformed into mel-scaled spectrograms. These spectrograms express human perceptible image representations of audio signals and serve as training input for the \textit{Deep Neural Autoencoder} provided by the \textit{auDeep} toolkit. It is trained to reduce the dimensions of the audio representations \cite{freitag2017audeep}. By applying the trained encoder component in our concept, the input data is compressed and the samples are reduced to vectors with 1024 dimensions.
These vectors are the input of the \textit{Fingerprinter}, which trains a \textit{K-Means} clustering model \cite{pedregosa2011scikit}.
The distribution of samples in each cluster per radio station is regarded as the station's fingerprint.
The last component shown in Figure \ref{fig:architecture} is the \textit{Recommender}, which recommends radio stations similar to a particular input station on basis of the Euclidean distance of their fingerprints. 
A small distance implies a comparably high similarity (see Section \ref{sec:recommender} for more details).

\subsection{Data Collector}
\label{chapter:data-collector}
The \textit{DataCollector} unit (see Figure \ref{fig:architecture}) consists of several components that collect and process radio data. One of them records radio stations by requesting a list of radio services and their HTTP \textit{bearers}\footnote{Connection information, how a radio station can be received by a device (broadcast or streaming)} from the \textit{HRADIO Communication Platform} ~\cite{friedrich2019distributed} via its REST API. Currently, a list of 461 valid and unique streams is received. The HTTP stream address of a sender is queried in order to receive the audio bytes of the corresponding audio stream using ICY ('I Can Yell') tag technology. The ICY tag protocol defines the transmission of textual content within audio streams and is commonly used due to its simple integration. 
Using \textit{RabbitMQ}\footnote{An open source message broker (\url{https://www.rabbitmq.com/})} the received audio bytes are transferred to a separate component which is responsible for extracting audio snippets out of the radio streams. 
During 24 hours, each station is recorded for five seconds within intervals of two minutes.
In order to reduce the amount of news sequences included in the recorded snippets, the five minutes before and after full hours are not recorded. 
This leads to a total number of 576 samples per radio station. Thereby, a full day of samples for 431 radio stations could be recorded, while 30 stations could not be recorded entirely due to server-side connection problems.
In total, we collected 266,239 audio snippets, whereas 17,983 belong to incomplete radio station recordings.

\subsection{Deep Fingerprints}
\label{sec:deepfinger}
\begin{figure}[!htbp]
\centering
\vspace{4mm}
\includegraphics[width=0.8\textwidth]{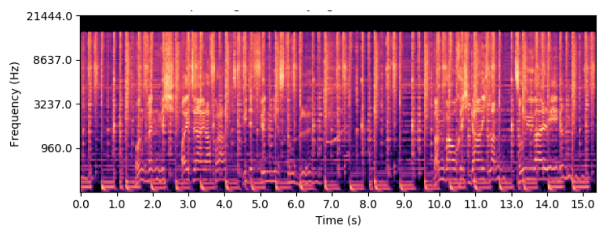}
\vspace{4mm}
\caption{An example of a mel spectrogram with 128 mel features. The x-axis represents the time scale and the y-axis the frequency scale.}
\vspace{1mm}
\label{fig:spectrogram}	
\end{figure}

The audio signals extracted in the previous step are transformed into a time-frequency representation, called mel spectrogram (see Figure \ref{fig:spectrogram}), using the \textit{auDeep} toolkit. Mel spectrograms are close to the human perception of audio signals with the mel scale being an assignment of actual to perceived frequencies.
In the example in Figure \ref{fig:spectrogram}, time is depicted on the x-axis and f frequencies are shown on the y-axis. 
In addition, the colour represents frequency amplitudes (or loudness): dark areas indicate low amplitudes, bright areas indicate high (or loud) amplitudes. For mel spectrogram generation we use the hyperparameters suggested by \textit{auDeep} (a mel scale of magnitude $128$, a window width of $0.08$ with an overlap of $0.04$, a fixed length of 5 seconds per input snippet, and clipping of values below $-60 db$) \cite{freitag2017audeep}. 

\begin{figure}[!th]
\centering	
	\begin{minipage}[c]{.5\textwidth}
	\includegraphics[width=1\textwidth]{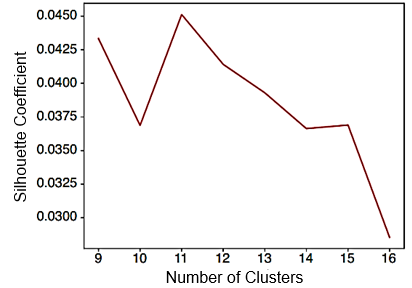}
	\end{minipage}%
	\begin{minipage}[c]{.5\textwidth}
  	\includegraphics[width=1\textwidth]{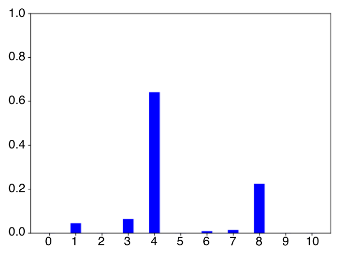}
	\end{minipage}
\caption{Silhouette coefficents for different numbers of clusters (left) with an optimum cluster count of 11. On the right side a sample histogram of \textit{BR KLASSIK} is depicted after the feature extraction process. The cluster index is depicted on the x-axis whereas the per-cluster frequency of occurrence is on the y-axis.}
\vspace{4mm}
\label{fig:silhouete}
\end{figure}

Subsequently, the \textit{auDeep} autoencoder was trained on all 266,239 audio files, whereas the fingerprinting is only applied to complete sets of audio snippets (for 431 radio stations).
The network is trained across 64 epochs with a batch size of 64 on 2 layers with each having 256 gated recurrent units (GRU), a learning rate of $0.001$ and a dropout of $0.2$.
Training the network for 7 days resulted in a loss of $0.237$.
After the training is finished the \textit{Fingerprinter} creates fingerprints of all complete stations, using all 266,239 vectors generated by the \textit{auDeep} component.
All these data points are divided into $n$ clusters by applying the \textit{K-Means} clustering algorithm.
The parameter $n$ is determined by using the silhouette coefficient \cite{aranganayagi2007clustering}, resulting in a range from $9$ to $16$ and showing a peak value at 11 as can be seen in Figure \ref{fig:silhouete}.
Each data point is assigned to exactly one cluster and the fingerprint is derived from a histogram across all clusters for each station.
This fingerprint vector now serves as input for the \textit{Recommender}.
\subsection{Recommender System}
\label{sec:recommender}
Based on learned per-station fingerprints (see Section \ref{sec:deepfinger}) it is possible to define a similarity metric for radio stations which is essentially the Euclidean distance between fingerprint vectors, considering the space of all $11$-dimensional fingerprint vectors as an Euclidean vector space. This similarity metric can be used to establish a content-based comparison of radio stations (the fingerprint represents the content of a station): A small distance between two station fingerprints implies high similarity.
Using this mechanism for recommendations, there are essentially two possibilities: 
\begin{enumerate}
    \item The $k$ nearest radio stations are suggested to the user as similar.
    \item Only radio stations within a certain Euclidean distance are listed.
\end{enumerate}
The first possibility carries the risk that the returned station list may also contain distant radio stations. In contrast to that, for the second advance, the distances of provided radio stations are small in any case, while it carries the risk that the result set may be empty.
\par
Because of the subjectivity of user taste and ratings, providing high-quality recommendations is not a trivial task and cannot only be based on station fingerprints. 
The \textit{HRADIO Communication Platform} offers therefore a multitude of different recommendation strategies with the fingerprint-based recommender being one of it.
The different modules are highly configurable and can be combined to a customized radio station recommendation experience. 
Examples for additional modules are trend-, metadata- and location-based recommenders.

\section{Evaluation}\label{sec:evaluation}
In this section, the recommender system is evaluated quantitatively and its results are discussed by example (Section \ref{sec:eval_rec}).
Furthermore, we were interested in identifying station archetypes that represent certain categories (or genres) of stations. For that purpose, an \textit{Archetypal Analysis} of the station fingerprint dataset was conducted with its results being presented in Section \ref{sec:archetypal_analysis}.
Finally, fingerprint deltas between different times of day are analyzed and discussed while highlighting some significant examples in Section \ref{sec:day_time}.

\subsection{Recommender System}
\label{sec:eval_rec}
The evaluation of the recommender system is based on the station fingerprint dataset and corresponding genre labels that stem from existing metadata as visualized in Figure \ref{fig:pca_plot}.
In the following, we show how the distance measures map to these genre labels by choosing the most meaningful examples.

Table \ref{tab:recommendations} shows the three closest radio stations (according to their fingerprint distance) with their distance to a requested service.
The station \textit{BR Klassik} is assigned to the genres \textit{Classical Music} and \textit{Special Music}.
The closest 3 stations are \textit{NDR Kultur} with the genres \textit{Classical Music} and \textit{Cultural} with an Euclidean distance of $59.58$, \textit{HR 2} with the genres \textit{Classical Musik}, \textit{Cultural}, and \textit{Special Music} and an Euclidean distance of $60.81$, and \textit{Classic FM} with the genres \textit{Classical Music} and \textit{News} and an Euclidean distance of $60.93$.
\begin{table*}[!htbp]
\vspace{4mm}
\centering
\resizebox{\textwidth}{!}{%
\begin{tabular}{|l|l|l|l|}
\hline
\textbf{Requested station} & 1st closest station  & 2nd closest station & 3rd closest station \\ \hline
\textbf{\begin{tabular}[c]{@{}l@{}}BR Klassik\\ \textit{Classical Music}\\ \textit{Special Music}\end{tabular}} & \begin{tabular}[c]{@{}l@{}}NDR Kultur\\ \textit{Classical Music}\\ \textit{Cultural}\\ distance: 59.58\end{tabular} & \begin{tabular}[c]{@{}l@{}}HR2\\ \textit{Classical Music}\\ \textit{Cultural}\\ distance: 60.81\end{tabular}    & \begin{tabular}[c]{@{}l@{}}Classic FM\\ \textit{Classical Music}\\ \textit{News}\\ distance: 60.93\end{tabular} \\ \hline

\textbf{\begin{tabular}[c]{@{}l@{}}Heart UK\\ \textit{Classic/Dance/Pop-rock}\\ \textit{Disco}\end{tabular}} & \begin{tabular}[c]{@{}l@{}}Capital XTRA Reloaded\\ \textit{Rap/HipHop/Raggae}\\ distance: 41.09\end{tabular} & \begin{tabular}[c]{@{}l@{}}3FM Isle of Man\\ \textit{Hit-Chart}\\ distance: 86.34\end{tabular} & \begin{tabular}[c]{@{}l@{}}FFH Rock\\ \textit{Rock}\\ \textit{Soft Rock}\\ distance: 88.29\end{tabular} \\ \hline

\textbf{\begin{tabular}[c]{@{}l@{}}LBC UK\\ \textit{Local/Regional}\\ \textit{News}\end{tabular}}  & \begin{tabular}[c]{@{}l@{}}Bayern 5 Plus\\ \textit{Information}\\ distance: 117.97\end{tabular} & \begin{tabular}[c]{@{}l@{}}WDR 3\\ \textit{Classical Music}\\ \textit{Cultural}\\ distance: 118.22\end{tabular} & \begin{tabular}[c]{@{}l@{}}SWR 2 Archiv Radio\\ \textit{Documentary}\\ distance: 118.25\end{tabular} \\ \hline
\end{tabular}%
}
\vspace{2mm}
\caption{Three examples of recommendation requests and the three closest results including their Euclidean distances.}
\label{tab:recommendations}
\vspace{4mm}
\end{table*}

\par
All 3 stations near to \textit{BR Klassik} play similar content within the genres \textit{Classical Music} and \textit{Cultural}.
The second radio station \textit{Heart UK} is assigned to the genres \textit{Classic/Dance/Pop-rock}, \textit{Disco}, \textit{Local/Regional}, \textit{Dance/Dance-pop}, and \textit{Showbiz}.
The closest three stations are \textit{Capital XTRA Reloaded} with the genres \textit{Rap/Hip Hop/Reggae} with an Euclidean distance of $41.09$, \textit{3FM Isle of Man} with the genre \textit{Hit-Chart} and an Euclidean distance of $86.34$, and \textit{FFH Rock} with the genres \textit{Rock}, \textit{Soft Rock}, \textit{Grunge}, \textit{Heavy Rock}, and \textit{Rock \& Roll} and an Euclidean distance of $88.29$.

\par
The only station close to \textit{Heart UK} by genre is \textit{3FM Isle of Man}.
The other two stations can be considered not similar, being assigned to \textit{Rap/HipHop/Raggae} versus \textit{Rock, Soft Rock, Grunge, Heavy Rock, Rock \& Roll}.
The third radio station \textit{LBC UK} is assigned the genres \textit{Non-fiction}, \textit{Local/Regional}, and \textit{News}.
The closest 3 non-variant stations are \textit{Bayern 5 Plus} with the genre \textit{Information} with an Euclidean distance of $117.97$, \textit{WDR 3} with the genres \textit{Classical Musik} and \textit{Cultural} with an Euclidean distance of $118.22$, and \textit{SWR 2 Archiv Radio} with the genre \textit{Documentary} and an euclidean distance of $118.25$.
\par
All three stations near to \textit{LBC UK} publish a lot of spoken content, as suggested by their genres \textit{Local/Regional}, \textit{News}, or \textit{Documentary}.

\par
As we could show and explain by example, the proposed station fingerprints in combination with the discussed similarity metric is a well-working approach for content-based station recommendations which outputs similar stations for a given station.  

\subsection{Archetypal Analysis}
\label{sec:archetypal_analysis}
\begin{figure}[!htbp]
\centering
\vspace{4mm}
\includegraphics[width=0.8\columnwidth]{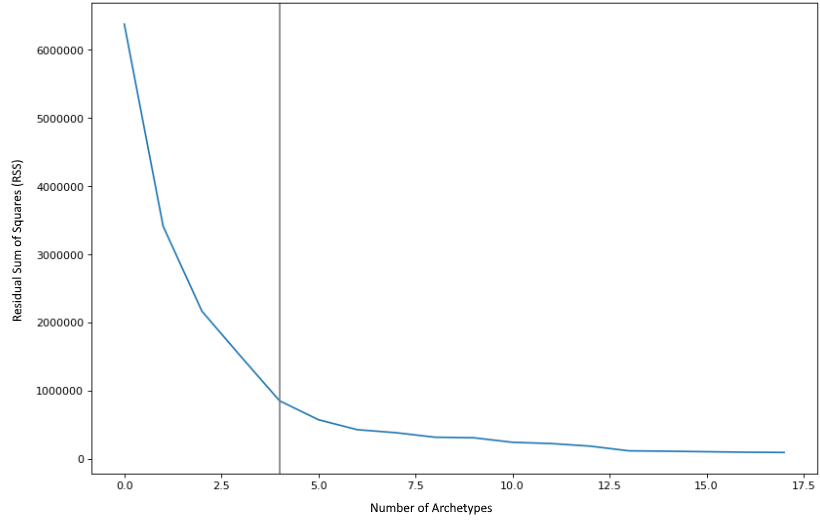}
\vspace{1mm}
\caption{Number of archetypes and corresponding RSS values. This so-called scree plot is used for the selection of the best number of archetypes.}
\vspace{2mm}
\label{fig:elbow}	
\end{figure}

In order to distill station categories, we conducted an \textit{Archetypal Analysis} on the fingerprints which is an unsupervised learning method used to extract representative individuals in a dataset. 
A data point is defined by its affiliation to \textit{k} archetypes ~\cite{cutler1994archetypal}. 
In soccer, for example, a player could be described by 10\% defender, 50\% midfielder, and 40\% striker.
We use this concept to find representative stations on basis of their fingerprints. 
In our case, the optimal amount of four archetypes was determined using the so-called \textit{elbow criterion} \cite{eugster2009spider}.
Therfore, the residual sum of squares (RSS) for different numbers of archetypes is visualized in a scree plot (see Figure \ref{fig:elbow}).
The optimal number of archetypes is the one where the curve has its strongest bend. 

\par
\begin{figure}[!htbp]
\centering
\vspace{4mm}
\includegraphics[width=0.8\columnwidth]{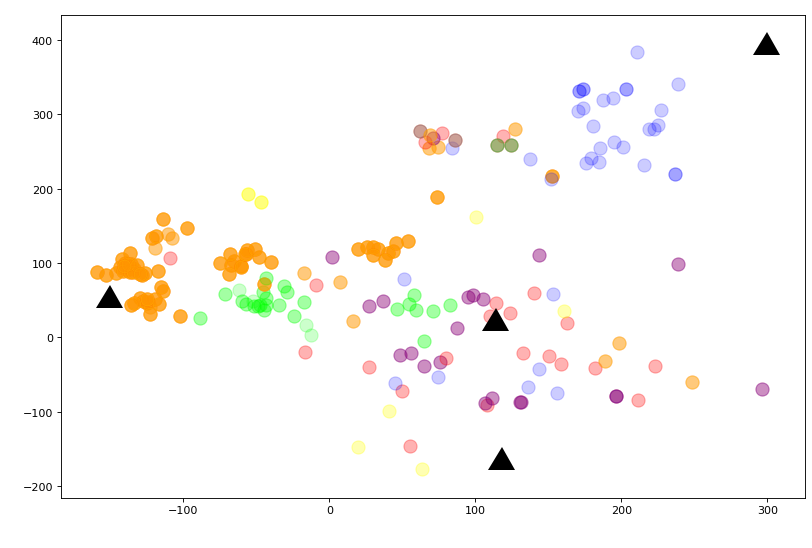}
\vspace{1mm}
\caption{Visualization of fingerprint vectors, using PCA. The dots mark the radio services, colored by genre. Black triangles represent archetypes.}
\vspace{2mm}
\label{fig:pca_plot}
\end{figure}

Figure \ref{fig:pca_plot} shows a plot of all fingerprints, reduced to 2 dimensions by using a \textit{Principal Component Analysis (PCA)} \cite{jolliffe2011principal}, where stations are points coloured according to their genre.
Archetypes are represented as black triangles.
The first archetype is at $[115.77$, $-164.82]$. \textit{Antenne P} defined as genre \textit{Oldies} is the closest station to it with a distance of $19.86$.
Within a radius of an Euclidean distance of 150, 71 stations could be found in total.

\textit{SWR1 RP} is nearby next with a distance of $51.90$, encompassing the genres \textit{Pop Music} and \textit{Regional}.
The second archetype is at $[292.96$, $385.21]$. \textit{Noods Radio} is the closest station with a distance of $19.23$ not listing any genres.
Within a radius of an Euclidean distance of 150, 7 stations could be found in total.
\textit{95.3 KGY Olympia} is the next nearest station with a distance of $41.56$, classified as \textit{AOR / Slow Rock / Soft Rock}.
\par
The third archetype is located at $[-144.58$, $50.05]$. \textit{Heart Crawly} has a distance of $8.76$ to it, encompassing the genres \textit{Classic/Dance/Pop-rock}, \textit{Disco, Local/Regional, Dance/Dance-pop}, and \textit{Showbiz}.
Within a radius of an Euclidean distance of 150, 175 stations could be found in total.
The next other station providing genre metadata is \textit{Heart Dorset} with a distance of $11.32$, classified by the same genres.
\par
The last archetype is at $[138.10$, $60.20]$. \textit{Bayern 2 Nord} is the closest station with a distance of 88.83 and genre \textit{Cultural}.
Within a radius of an Euclidean distance of 150, 48 stations could be found in total.
The next other station offering genre meta data is \textit{Classic FM} with a distance of 99.47 to the archetype, playing \textit{Classical Music} and \textit{News}.
\par
In summary, the four found archetypes map well on genres and also target groups: The first archetype corresponds to stations with a target group that likes to listen to oldies / old pop music and demands for regional information. The second archetype has nearby stations with a very diverse programme that does not fit into specific music genres (\textit{Noods Radio} for example plays almost everything from post punk to electro) and thus has a very broad target group. The third archetype represents stations with a focus on dance music whereas the fourth archetype is related to classical music and news. 
Interestingly, there is no specific archetype for contemporary pop music which could be due to the fact that this kind of music is more or less a mixture of existing genres. 

\subsubsection{Comparing Recommendations by Day Times}
\label{sec:day_time}
It is well known that radio stations change their programme over the day with, for example, fast wake-up music in the morning and slower, relaxing music in the evening. 
Our hypothesis is, that this effect also manifests in the fingerprints of a station extracted for different day times. 
\par
\begin{figure}[!htbp]
\centering
\vspace{4mm}
\includegraphics[width=0.8\columnwidth]{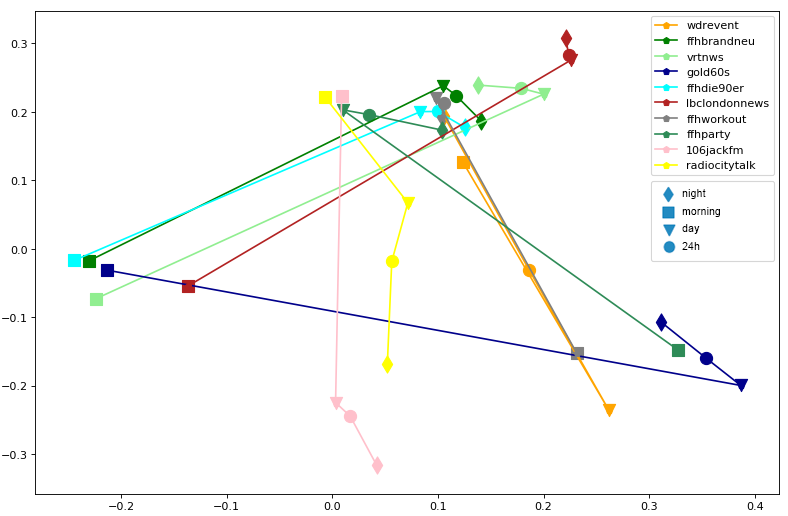}
\vspace{1mm}
\caption{Fingerprint trajectories visualizing deltas for stations for different times of day. Trajectory color corresponds to the station, circle shapes are whole-day fingerprints (24h), diamonds night, squares morning and triangles day time. }
\vspace{2mm}
\label{fig:finger_daytime}
\end{figure}

In order to test this hypothesis, we created three additional, normalized fingerprints per times of day, namely night, morning, and day.
Night fingerprints consider audio samples from 09:00 pm to 05:00 am, morning fingerprints consider samples from 05:00 am to 09:00 am, and day fingerprints consider samples from between 09:00 am and 09:00 pm.
See Figure \ref{fig:finger_daytime} for a visualization of the different fingerprints using per-station trajectories.
Comparing those three fingerprints to each other and to the whole-day fingerprints provides interesting insights into how radio stations change throughout the day.
\par
E.g., large changes appear on \textit{Gold 60s} which had comparably large distances between morning (square) and other times of the day (triangle, diamond) and whole day (circle)  fingerprints, leading to the assumption that the style of content changes heavily between morning and rest of the day.
This pattern is also visible in other stations' trajectories (e.g. \textit{FFH Die 90er}, \textit{FFH BrandNeu} and \textit{106 Jack FM}).
On the other hand, some stations show only minor (\textit{FFH Workout}) or moderate (\textit{Radio City Talk}) changes over the day.
\par
\begin{figure}[!htbp]
\centering
\vspace{4mm}
\includegraphics[width=0.8\columnwidth]{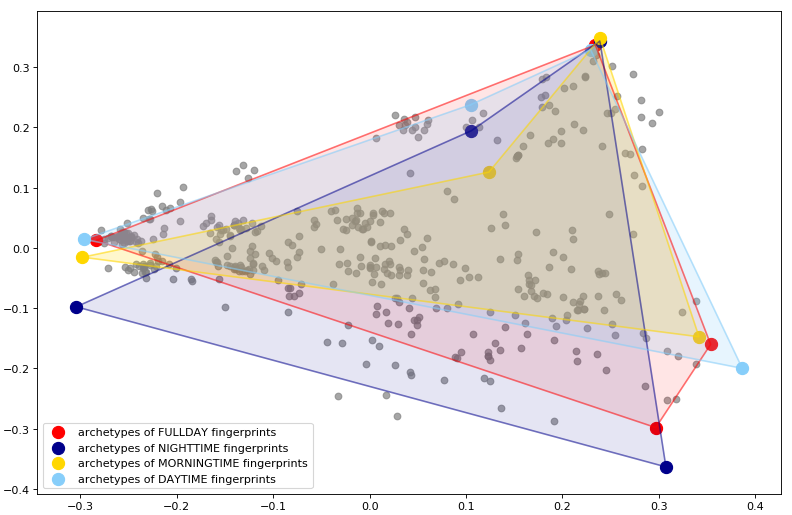}
\vspace{1mm}
\caption{Extracted archetypes for different times of day.}
\vspace{2mm}
\label{fig:arche_daytime}
\end{figure}

An additional hint for daytime-based changes visible in the extracted fingerprints is depicted in Figure \ref{fig:arche_daytime}. 
The day time and morning time archetypes (light blue, yellow) significantly differ from the night time archetypes (dark blue) which clearly shows that the assumed programme difference dependent on the time of day is visible in the fingerprint data-set as well.
\par 
To summarize, we could prove our hypothesis, that time dependent radio content differences are visible in the fingerprint data by showing exemplary per-station differences and differences in the shapes and locations of the archetypes. 
\section{Conclusion}\label{sec:conclusion}
In this paper, we presented a holistic pipeline for deep radio station fingerprinting and recommendation.
The first component utilized throughout the process is the \textit{HRADIO Communication Platform}, which delivers metadata and bearer addresses of radio stations.
The next component, the \textit{DataCollector} recorded 461 radio stations over 24 hours, generating a total of 266,239 audio samples. Following this step, we trained a \textit{Deep Neural Autoencoder} with \textit{auDeep} in order to reduce each sample's dimensions.
The \textit{Fingerprinter} then clustered the samples and created a deep fingerprint for each service.
The last component, the \textit{Recommender}, is able to give recommendations depending on the calculated distance between the fingerprints.
\par
We evaluated the \textit{Recommender} by analyzing the recommendation results and noticed similar genres for close radio stations.
Additionally, we conducted an \textit{Archetypal Analysis} on the fingerprints, leading to four archetypes.
By analyzing their closest radio stations, we noticed dissimilar target groups and genres between the archetypes and similar ones between the closest stations surrounding them.
\par
Finally, we compared whole-day fingerprints to fingerprints extracted in the night, morning and day time.
The current time of day seems to have a big influence on recommendations for some services. 
In contrast to that, others stayed somewhat constant throughout 24 hours.
The data-set we created offers many more opportunities for further research and analysis.
To further validate this, user studies should follow with which the quality of a recommendation can be quantified.
In addition, the night-morning-day approach could be evaluated in far more depth.
\par 
We see three important practical use cases for the aforementioned techniques and findings:
\begin{itemize}
    \item The proposed content-based recommender can complement existing recommender engines that mostly use collaborative filtering which is rather domain-unspecific. Within the HRADIO project, a station recommender system was developed, that uses a fingerprint-based recommendation module together with modules based on metadata, location and user behavior.\footnote{Android App that uses the modularized station recommender: \url{https://play.google.com/store/apps/details?id=lmu.hradio.hradioshowcase}}
    \item The described audio analysis pipeline can be used to automatically extract station and programme metadata (e.g. genre-like classifications) which is a big step forward considering the current situation regarding metadata availability in the radio domain.
    \item The extracted fingerprints and the proposed distance metric can also be helpful for radio stations that would like to check if their current play-out matches with the station's target programme and groups. This check can be conducted automatically and could provide even more detailed guidance like what to play in the next two hours to match the taste of a certain target group.
\end{itemize}
\par
In summary, this approach provides valuable recommendations only on basis of audio signals and without the need of additional metadata. 
Moreover, the presented concepts are included and available within the open \textit{HRADIO Communication Platform} and thereby enable also small businesses and radio stations to actively take part in a radio service landscape driven by smart services. 

\bibliography{refs} 
\end{document}